\RequirePackage{fix-cm}
\documentclass[smallextended]{svjour3}       
\smartqed  
\usepackage{graphicx}
\usepackage{amsmath}
\usepackage[colorinlistoftodos, color=green!40, prependcaption]{todonotes}

\begin{document}
\title{Relativistic rigid systems and the cosmic expansion}



\author{Luciano Combi*         \and
        Gustavo E. Romero 
}


\institute{L. Combi* \at
              Instituto Argentino de Radioastronom\'ia, C.C. No. 5, 1894, Villa Elisa, Argentina \\
              Tel.: +54-45-678910\\
              \email{lcombi@iar.unlp.edu.ar}           
           \and
           G.E. Romero \at
              Instituto Argentino de Radioastronom\'ia, C.C. No. 5, 1894, Villa Elisa, Argentina, and \\
              Facultad de Ciencias Astron\'omicas y Geof\'isicas, Universidad Nacional de La Plata, Paseo del Bosque s/n, 1900 La Plata, Buenos Aires, Argentina.         
}

\date{Received: date / Accepted: date}

\maketitle

\begin{abstract}
We analyze the necessary conditions for a body to remain rigid in an expanding cosmological Universe. First, we establish the main theorems and definitions for having a rigid body in a general spacetime as well as the new concept of quasilocal rigidity. We apply the obtained results to a homogeneous universe exploring the differences with flat spacetime. We discuss how the concept of rigid body helps to understand the expansion of space in cosmology. Finally, using a rigid system as a reference frame, we calculate the gravitational energy, and we compare it with previous results in the literature.

\end{abstract}

\section{Introduction}

A rigid body is a physical system that cannot be deformed, i.e. a body in which distances between its parts are constant. In Newtonian physics, rigidity is well-defined and essentially non-local; for instance, if we push a rigid body, each part must move in a way that that distances remain invariant. This implies that the interaction between all parts must be instantaneous. In Relativity, this is prohibited by the causal structure of spacetime, i.e. nothing can move faster than light.

In 1909, a few years after Einstein's presentation of Special Relativity, Born proposed the first definition of relativistic rigidity \cite{born}. He showed that a Lorentz invariant concept of rigid motion is possible to formulate \footnote{As Pauli remarks, it is natural to introduce the concept of rigid motion, while the concept of rigid body does not have place in Relativity. In this paper, when we refer to a rigid body, we mean a body that mantains rigid motion.}, but contrary to Newtonian physics, each part of the body has to accelerate at different rates to stay rigid. That same year, Paul Ehrenfest realized that the intrinsic geometry of a rotating rigid body (in the sense of Born) is not Euclidean \cite{ehrenfest}. This \textit{gedankenexperiment} was crucial for Einstein's development of General Relativity, where spacetime itself is curved \cite{stachel}.

In the presence of curvature, a rigid system has to accelerate to avoid tidal deformations. The rigidity condition, however, cannot always be maintained, e.g. if a body enters a black hole, it will be unavoidably deformed. On the other hand, the Universe as a whole is itself curved, in accelerated expansion. Distances between free bodies, such as galaxies, will naturally expand with the Universe. This leads us to a fundamental question: can we build rigid bodies if space itself is expanding? Moreover, does the answer depends on the universe model? Is it possible to hold a rigid motion eternally? 

In this work, we provide a detailed answer to these questions. This will allow us to analyze some fundamental features of cosmology and the nature of expansion. he paper is organized as follows: In Section 2, we present the concepts of rigidity that we adopt throughout the paper, the one by Born and a recently introduced quasilocal notion of rigidity. In Section 3, we give a fresh look at the nature of rigidity in flat spacetime, giving a general prescription to build rigid systems and represent them. In Section 4, we discuss the expansion of space in cosmology. We analyze what kind of rigid systems can be built in this scenario. Finally, in Section 5, we apply the results of the previous section to calculate the gravitational energy of a compact region of the universe. 

\section{Rigid systems in relativity}
\label{sec: rigidity}

\subsection{Born-rigidity}
Let us consider a relativistic system represented by a congruence $\mathcal{C}$ defined as a tube of time-like world-lines. Given a spacetime model $( g_{ab}, \mathcal{M})$, the congruence has a unit time-like vector field $u^a(x)$ tangent to the world-lines. This vector allows to define a local space 3-metric, $h_{a b}:= g_{a b} + u_{a} u_{b}$ in $\mathcal{C}$. In order to analyze how the system evolves in \textit{space}, let us consider a connecting vector $\xi^a$ in the congruence, lying in the orthonormal space of $u^a$, i.e $h^{a}_b \xi^{b}= \xi^{a}$. By definition, this vector $\xi^a$ is Lie-transported along the congruence:
\begin{equation}
    (\mathcal{L}_{u}\xi)^{a}= u^{b}\nabla_{b} \xi^{a} - \xi^{b} \nabla_{a} u^{b} =0.
\label{eq: liexi}
\end{equation}

We define now a proper space-distance within the system as
\begin{equation}
l^2:= g_{ab} \xi^{a} \xi^{b} \equiv h_{ab} \xi^{a} \xi^{b},
\label{eq: length}
\end{equation}
i.e. the square length of a connecting vector in the rest frame of the congruence \cite{felice}. Next, let us split the covariant derivative of $u^a$ as:
\begin{equation}
    \nabla_{b} u_{a}= h_{a}^{c}h_{b}^{d}\nabla_{d} u_c - \alpha_{a}u_{b}= \Theta_{ab}+\omega_{ab}-\alpha_{a}u_{b},
    \label{eq: covu}
\end{equation}
where $\omega_{ab}:= h^{a'}_{a} h^{b'}_{b}\nabla_{[b'}u_{a']}$ is the vorticity, $\Theta_{ab}:=h^{a'}_{a} h^{b'}_{b} \nabla_{(a'} u_{b')}$ is the expansion tensor, and $\alpha^a:=u^b \nabla_b u^a$ is the acceleration of the congruence. We can further decompose the expansion tensor $\Theta_{ab}$ in its trace irreducible parts as $\Theta_{ab}= \bar{\Theta}_{ab} + \frac{1}{3} \Theta h_{ab}$, where $\bar{\Theta}_{ab}$ is the shear tensor and $\Theta:= h^{a}_{b} \nabla_a u^b \equiv \nabla_a u^a$  is the expansion scalar (see e.g. Ref. \cite{rezzolla}). The evolution of the system's shape is given by the time derivative of the distance $l$ in Equation (\ref{eq: length}). Using Equations (\ref{eq: liexi}) and (\ref{eq: covu}), we obtain that the space length evolution is determined by:
\begin{equation}
   u^a \nabla_a l^2 = 2\Big(\bar{\Theta}_{ab} + \frac{1}{3} \Theta h_{ab}\Big) \xi^{a} \xi^{b},
\end{equation}
which implies, in turn, that the relative change of distances is given by:
\begin{equation}
   (u^a \nabla_a l)/l = \bar{\Theta}_{ab} e^{a} e^{b} + \frac{1}{3} \Theta,
   \label{eq: ldot}
\end{equation}
where $e^a$ is the normalized $\xi^a$. Thus, if the proper space length of a body does not change in time, it implies that both $\bar{\Theta}_{ab}$ and $\Theta$ must vanish, or equivalently $\Theta_{ab}=0$, i.e. there are no deformation nor expansions in the object. We define then the following concept of rigidity,  called \textit{Born-rigidity} \cite{born}.

\begin{description}
\item[\textbf{Born rigidity}]: A physical system represented by a congruence $\mathcal{C}$ with tetravelocity $u^a$ in an arbitrary spacetime model $( \mathcal{M},  g_{ab})$, is Born-rigid if the spatial distances orthogonal to $u$ do not change along the congruence, i.e. if $u^a \nabla_a l=0$ in Equation (\ref{eq: ldot}) .
\end{description}

For a Born-rigid physical body, the following conditions are equivalent (c.f. Refs. \cite{pirani} and \cite{giulini}):
	\begin{itemize}
	\item[(A)] The expansion tensor of the congruence is zero, $\Theta_{ab}(u^c)=0$, which immediately implies that the expansion and the shear are zero, $\Theta(u^c)=0$ and $\bar{\Theta}_{ab}(u^c)=0$.
	\item[(B)] The Lie derivative along the congruence of the locally defined space metric in each point of the body is zero, $ (\mathcal{L}_u h)_{ab}= 0$.
	\item[(C)] In an adapted coordinate system, $\lbrace T, X^i \rbrace$, the space metric is independent of time, $\partial_T h_{ij}=0$.
	\item[(D)] The covariant derivative of the tetravelocity is determined by the vorticity and the acceleration as
	\begin{equation}
	\nabla_{b} u_{a}= \omega_{ab} - \alpha_{a} u_b,
	\label{eq: rigideq}
	\end{equation}
or, directly taking the anti-commutator in Eq.  \eqref{eq: covu} and using definition (A):
	\begin{equation}
	\alpha_{(a}u_{b)} +\nabla_{(b}u_{a)} =0.
	\label{eq: rigideq}
	\end{equation}
\end{itemize}

The Born-rigidity condition thus imposes 6 independent partial differential equations to the 3 components of the tetravelocity of the congruence. Because of the dependence of $h_{ab}$ with $u_a$, the equations are non-linear. Since this is an over-constrained system, i.e. 6 constrains for 3 components, integrability conditions can be derived. Following Pirani and Williams \cite{pirani}, taking $u^a$, $\omega_{ab}$, and $\alpha_a$ as variables, one can derive a set of equations for their first derivatives in terms of algebraic combinations of these variables. 

In Newtonian physics, as it is well-known, a rigid body has six degrees of freedom, meaning that we can maintain a body in a rigid state under any translations or rotations \cite{epp1}. In a relativistic framework, since we lack an absolute frame where all events are simultaneous, rigid motions are limited by the causal structure of spacetime; this is reflected by the over-constrained equations of a Born-rigid body. Indeed, for some transformations, maintaining the rigidity of an extended body may require super-luminal velocities of some of its parts. The type of transformations that maintain rigidity depends also on the background spacetime. 

In a spacetime with symmetries, there are preferred directions, characterized by the Killing vectors of the metric. A time-like Killing vector $T^a$ generates a preferred congruence with tetravelocity $u^{a}:= T^a ||T||^{-1}$. The congruence is then said to follow an isometric or \textit{Killing motion}. There is a direct connection between symmetries and rigid motions of a congruence given by the following known theorem:

\begin{description}
\item[\textbf{Theorem} I.] Any congruence $\mathcal{C}$ in a general spacetime following a Killing flow $u^a$ is Born-rigid (see Ref. \cite{pirani}).

\item[Proof.] If $u^a$ is a Killing motion, then $u^a= T^a ||T||^{-1}$, where $T^a$ is a Killing vector and $||T||$ its normalization. We thus have that the Killing covector is $T_a \equiv ||T|| u_a$. Using this expression in the Killing equations for $T_a$ and developing the derivatives we have:
\begin{equation}
\Big(\nabla_{(a}||T||\Big) u_{b)} + ||T|| \nabla_{(b}u_{a)}=0,
\label{torigid}
\end{equation}
and if we contract with $u^b$, we obtain:
\begin{equation}
(u^b\nabla_b ||T||) u_a - \nabla_a ||T|| + ||T|| \alpha_a=0.
\label{accelT}
\end{equation}
Contracting again with $u^a$ we get $u^b\nabla_b ||T||\equiv 0$, which implies from Eq. \eqref{accelT}:
\begin{equation}
\alpha_a = \frac{\nabla_a ||T||}{||T||}.
\label{accelscalar}
\end{equation}

Inserting this in Eq. \eqref{torigid}, we have:
\begin{equation}
||T|| \Big( \alpha_{(a}u_{b)} +\nabla_{(b}u_{a)} \Big)=0,
\end{equation}
and since $||T||\neq0$, this means that the Killing flow obeys the Born-rigidity equations (see definition (D)).
\end{description}

An immediate important corollary follows from this theorem.

\begin{description}
\item[\textbf{Corollary I}.] Given a rigid congruence in an arbitrary spacetime with tetravelocity $u^a$, the congruence is a Killing flow if and only if the following relation holds:
\begin{equation}
\nabla_{[a} \alpha_{b]} =0.
\label{killingacc}
\end{equation}

\item[Proof.] If the rigid congruence is a Killing flow, the tetravelocity obeys Eq. \eqref{accelscalar} from Theorem I, and thus $\alpha_a = \nabla_a \Phi$, where we define the scalar $\Phi:=\log(||T||)$. In this way Eq. \eqref{killingacc} follows easily. On the other hand, if we have a rigid congruence with a tetravelocity obeying $\nabla_{[a} \alpha_{b]} =0$, then its acceleration is an exact differential, which again means that $\alpha_a = \nabla_a \Phi$. With this, and the rigidity condition $\alpha_{(a}u_{b)} +\nabla_{(b}u_{a)} =0$, we recover the Killing equation following the inverse path from the derivation in Theorem I.

\end{description}
\textit{Remarks:} The linear combination of Killing vectors is also a Killing vector. From Theorem I, given two Killings vectors $T^a_{(1)}$ and $T^a_{(2)}$, the tetravelocity $u^{a}:= (\alpha T^a_{(1)}+\beta T^a_{(2)})||\alpha T^a_{(1)}+\beta T^a_{(2)}||^{-1}$ represents a rigid motion if $u^a$ is time-like. Note that in general a Killing motion is not geodesic, even in Minkowski spacetime (see Section 2.2). 

On the other hand, from Eq. (\ref{eq: rigideq}), rigid motions can be classified as rotational, $\omega_{ab} \neq 0$, or irrotational, $\omega_{ab} = 0$. Rotational motion is heavily constrained by the integrability conditions (c.f. Equations (4.2) and (4.3) in \cite{pirani}). In particular, for maximally symmetric spacetimes in four dimensions, it is possible to show the following lemma:

\begin{description}
\item[\textbf{Lemma} I] Given a Born-rigid body with tetravelocity $u^a$ in a maximally symmetric spacetime, the Lie derivative of the vorticity is zero:
\begin{equation}
\mathcal{L}_u \omega_{ab} =0,
\end{equation}

\item[Proof.] We generalize the proof from Ref. \cite{pirani} to maximally symmetric spacetime. Let us consider the Lie derivative of the Levi-Civita connection in the direction of a time-like vector field $u^a$. Using the Lie Derivative definition, by direct computation, we have:
\begin{equation}
\mathcal{L}_u \Gamma^{c}_{ab} = \nabla_b \nabla_a u^c + R^{c}_{abk} u^k.
\end{equation}

Taking the commutator of this expression, it is easy to show that:
\begin{equation}
2 \nabla_{[d} \mathcal{L}_u \Gamma^{c}_{a]b} = \mathcal{L}_u R^{c}_{bad}.
\label{lievort}
\end{equation}

Now, note that the from the definition of the vorticity:
\begin{equation}
\omega_{ab}:=  h^{a'}_{a} h^{b'}_{b}\nabla_{[b'}u_{a']} =  h^{a'}_{a} h^{b'}_{b}  \Gamma^{k}_{[a'b']} u_k,
\end{equation}
so applying twice the projector $h^a_b$ onto Equation \eqref{lievort} and considering the rigidity condition $\mathcal{L}_u h^{a}_{b}=0$, we get:
\begin{equation}
\mathcal{L}_u \Big( \omega_{ab} \omega_{cd} + {}_{\perp} R_{abcd} \Big)=0,
\end{equation}
where ${}_{\perp} R_{abcd}$ is the projected Riemann tensor on $h_{ab}$. If we consider a maximally symmetric spacetime, then 
\begin{equation}
R_{abcd}\equiv C_0 (g_{ab}g_{cd} -g_{ac}g_{bd}),
\end{equation}
with $C_0$ a constant, so we have that $\mathcal{L}_u {}_{\perp} R_{abcd} =0$ for rigid motions (see definition (B)). This implies, contracting again with the (antisymmetric) vorticity, that:
\begin{equation}
\mathcal{L}_u  \omega_{ab} =0.
\end{equation}
\end{description}

\textit{Remarks:} This lemma implies that we cannot set to rotation a resting rigid body without deforming it; this is sometimes called the Eherenfest paradox. Because of this, it is usually stated that a relativistic rigid body in flat spacetime has only three degrees of freedom instead of six as the Newtonian case. An important theorem regarding this kind of rotational rigid motions possible in Minkowski spacetime was given by Herglotz and Noether (see Ref. \cite{herglotz} and \cite{noether}), who proved that every rotational rigid motion in Minkowski is necessarily a Killing motion. We can generalize this theorem to a maximally symmetric spacetime:

\begin{description}
\item[\textbf{Theorem} II] \textit{(Herglotz-Noether)} A congruence following a rotational rigid motion in a maximally symmetric spacetime is a Killing motion.

\item[\textbf{Proof}] Using the definition of the Riemann tensor,
\begin{equation}
\nabla_a \nabla_b u_c -\nabla_b \nabla_b u_c = R^{k}_{cba} u_k,
\end{equation}
projecting onto $u^a$ and taking the antisymmetric part, we have explicitly:
\begin{equation}
u^c \nabla_c \omega_{ab}= u^c \nabla_c ( \alpha_{[a} u_{b]} ) + \nabla_{[b} \alpha_{a]} - \alpha^c \omega_{c[a} u_{b]}.
\end{equation}

Now, multiplying with $\omega^{ab}$, using Lemma I, and the antisymmetry of the vorticity, we have:
\begin{equation}
\omega^{ab} \nabla_{[b} \alpha_{a]} =0.
\end{equation}

This implies that a rotational rigid motion has, in general, $\nabla_{[b} \alpha_{a]} =0$, so by Corollary I, this implies that the rotational rigid motion is a Killing motion.
\end{description}

Note that in arbitrary spacetimes, rotational rigid motions could exist without following a Killing flow. From these theorems, we see that the notion of rigidity in relativity is quite restricted. This has motivated the search of more general definitions for a rigid body. Recently, a quasilocal definition of rigidity was proposed by Epp, Mann and McGrath \cite{epp1} in which the relativistic system recovers the six degrees of freedom of the Newtonian definition. This new concept of relativistic rigidity is also akin to quasilocal conservation laws for the gravitational energy, as it is a convenient reference frame to measure fluxes. In the next section, we briefly discuss this new notion of rigidity.

\subsection{Quasilocal rigidity}

Let us consider a congruence $\mathcal{C}$ defined as a family of time-like world-lines with topology $S^2 \times \mathbf{R} $. This means that the congruence $\mathcal{C}$ is a closed shell evolving in time. Given a spacetime model with a metric $g_{ab}$ defined over a manifold $\mathcal{M}$, the congruence has a natural time-like unit vector field, $u^a$ and an outward-directed unit space-like vector field $s^a$. The tetravelocity induces, as we saw, a spatial metric on the congruence, $h_{ab}$. As we assumed an $S^2$ space topology, i.e. the world-lines bounds a finite space region, we can use the orthonormal vector to the 2-sphere, and induce another spatial metric given by $\sigma^{ab}:= h^{ab}-s^a s^b$, which is well defined over the closed surface (see Ref. \cite{epp1}). In the same fashion, the spatial vector $s^a$ also induces a Lorentzian metric of a time-like 1+2 sheet moving on spacetime, defined as $\gamma^{ab}:=g^{ab}-s^a s^b$.

The evolution of the closed surface along the congruence might be characterized by kinematical quantities analogous to the ones we defined before. In particular, the surface expansion is characterized by the  expansion:
\begin{equation}
\theta_{ab}:= \sigma_{a}^{c} \sigma_{b}^d \nabla_{(c} u_{d)},
\end{equation}
from where we define the expansion scalar and the shear. A body represented by a congruence $\mathcal{C}$ would be rigid if $\theta_{ab}=0$; we then have the concept of shell-rigidity or quasilocal rigidity:

\begin{description}
\item[\textbf{Quasi-local rigidity}]: A physical system represented by a two-parameter congruence $\mathcal{C}$ with S$^2\times R$ topology is quasilocal rigid if its expansion tensor is zero, $\theta_{ab}=0$. 
\end{description}

The physical interpretation is similar to the one we gave before: the distances in the bounding surface remain constant. This generalization of the Born-rigidity concept allows us to build more general rigid bodies which, in particular, have six degrees of freedom as their Newtonian counterparts. It was shown in Refs. \cite{epp1}, \cite{epp2}, \cite{epp3} that quasilocal rigid systems can rotate and accelerate in an arbitrary spacetimes without losing rigidity. On the other hand, quasilocal systems are useful to characterize spacetime quantities because of the holographic nature of spacetime (see Ref. \cite{freidel}). In Section \ref{sec: gravenergy} we shall take these quasilocal rigid bodies as a proper frame to calculate the energy and momentum of an expanding universe. 

In the next section, we discuss how these fundamental notions of rigidity are useful to understand some of the dynamics of physical bodies on spacetime.

\section{Spacetime dynamics and rigidity}

In dynamical spacetimes, distances between geodesic objects change, e.g. by the passing of a gravitational wave or in expanding Universes. Thus, rigidity is a very distinctive property to have in a general spacetime. On the other hand, rigid systems constitute a natural reference frame. For instance, any inertial frame in Minkowski is a rigid system. If the frame, however, is in an arbitrary state of motion, rigidity will hold if the conditions of Section \ref{sec: rigidity} are satisfied. As we show in Theorem I, spacetime symmetries allow to build preferred rigid frames with the associated Killing vectors. For instance, a Lorentz boost, e.g. a boost along the x-axis with an associated Killing vector $\xi_{(x)}= x \partial_t + t \partial_x$, will generate an \textit{accelerated} Killing motion defined by $u^a= \xi^a/||\xi^a||$. It can be shown that the congruence will follow an hyperbolic trajectory, with constant acceleration on the boost direction \cite{giulini}. 

Another way to build local rigid frames in flat spacetime is by using geodesic coordinates, or Fermi coordinates. Given a time-like curve $\gamma^a(\tau)$ with an arbitrary acceleration $\alpha^a$ and vorticity $\omega_{ab}$ in an arbitrary spacetime, let us consider the class of space-like geodesics orthogonal to $\gamma^a(\tau)$ for each event $\tau$. This class of geodesics forms locally an hypersurface, where we can define a unique coordinate system $\lbrace T, \vec{X}=X^{i} \rbrace$, around the world-line (see Ref. \cite{mashhoon} for the formal construction).

Fermi coordinates are the natural extension of Cartesian (parallel) coordinates to arbitrary observers in a general spacetime. They are used for many applications since they allow a simple interpretation of local measurements in a gravitational field, as they naturally represent clocks and rigid rods in a curved background (see Ref.  \cite{binifelice}). 

In Minkowski spacetime, the coordinate system associated with an arbitrary world-line is:
\begin{equation}
\begin{split}
ds^2 & = \eta_{ab} dx^a dx^b \\
     & = - \Big[ ( 1+ \vec{\alpha} \cdot \vec{X} )^2 - (\vec{\Omega} \wedge \vec{X})^2 \Big] dT^2 + 2 ( \vec{\Omega} \wedge \vec{X})^2 \cdot d\vec{X} dT + \delta_{ij} dX^i dX^j.
\end{split}
\label{eq: fermimetric}
\end{equation}

A stationary congruence composed of world-lines that remain at $\vec{X}=$constant, would be accelerating and rotating according with the kinematics of the central world-line. The tetravelocity of this congruence is thus:
\begin{equation}
u^a = \delta^a_0 (( 1+ \vec{\alpha} \cdot \vec{X} )^2 - (\vec{\Omega} \wedge \vec{X})^2)^{-1/2}.
\label{eq: fermiobs}
\end{equation}

Since geodesic distances do not change in flat spacetime, any congruence composed of curves on a sphere of areal radius $||\vec{X}||=R_0$ will be rigid if the central geodesic itself does not change. In this way, we can build any rigid system with six constant parameters, given by $(\vec{\alpha}, \vec{\Omega})$ (see Ref. \cite{epp1}). If we first consider the case of a non-rotating system, $\vec{\Omega}=0$, the spacetime metric in these coordinates is:
\begin{equation}
ds^2= -( 1+ \vec{\alpha} \cdot \vec{X} )^2 dT^2 + d^3X.
\label{eq: acc}
\end{equation}

It is easy to show using Eqs. (\ref{eq: acc}) and (\ref{eq: fermiobs}) that the induced space metric, $h_{ab}= g_{ab} +u_a u_b$, does not change in time and it is in fact Euclidean, $h_{ab} \equiv \delta_{ab}$. Thus, a rigid accelerated frame with tetravelocity $u^a= \delta^a_0 ||(( 1+ \vec{\alpha} \cdot \vec{X} )||^{-1}$ suffers time dilatations among its parts, i.e. gravitational redshifts, but its local space remains Euclidean. These coordinates are the well-known Rindler coordinates in the M\"oller form \cite{misner} and correspond, as we saw above, to a boost Killing flow. Note that the system is well defined as long as $( 1+ \vec{\alpha} \cdot \vec{X} ) > 0$. Each part of the system, to maintain rigidity, must have different accelerations depending on the distance to the central system at $\vec{X}=0$ and the central acceleration $\vec{\alpha}$ \footnote{The parts of the body behind the direction of acceleration have to accelerate more than the parts ahead of it.}. The size of the system is thus limited by this central acceleration. For this frame, a horizon naturally appears, i.e. the Rindler horizon, limiting where the accelerated congruence can have zero expansion. This is relevant for the analysis of the Unruh effect in quantum field theory \cite{unruh}. Note that the appearance of a horizon arises from the rigidity condition. It is possible to show that constant accelerated congruences exist without this feature (see Ref. \cite{combi}). If the acceleration is time-dependent, we can see that the space metric remains unchanged, so rigidity is maintained even in that case.

In the case of a rigid rotating body, it is easy to show from (\ref{eq: fermimetric}) that the induced space metric is not flat due to the cross term $ 2 ( \vec{\Omega} \wedge \vec{X})^2 \cdot d\vec{X} dT $. Taking the tetravelocity (\ref{eq: fermiobs}), and a rotation velocity in the z-direction, $\vec{\Omega}= \Omega \partial_z$, the space metric of a rotating rigid system in cylindrical coordinates is:
\begin{equation}
h_{ij}dX^idX^j= dz^2 +d\rho^2 + \frac{\rho^2}{1-\rho^2 \Omega^2} d\phi^2,
\end{equation}
which is not Euclidean. This was first noted by Ehrenfest in the early years of Special Relativity \cite{ehrenfest}. If rotation is time-dependent, then the Lie derivative of the space metric is not zero and the system cannot maintain bulk rigidity (see Lemma I). As it is well-explained in Ref. \cite{epp1}, a relativistic system can rotate in a time-dependent way maintaining rigidity only on its ''shells´´, while the distances between these shells change. In the following section, we analyze how to apply all the previous concepts when spacetime itself is expanding.

\section{Space expansion in cosmology}

When spacetime is assumed to be homogeneous and isotropic, the most general solution of Einstein's field equations is the Friedmann-Lemaitre-Roberston-Walker (FLRW) metric, given in comoving coordinates by:
\begin{equation}
ds^2 = -dt^2 + a(t)^2 \Big( \frac{dr^2}{1-k r^2} + r^2d\Omega \Big),
\label{eq: flrwmetric}
\end{equation}
where $k\in \lbrace 0,+1,-1 \rbrace$ is the space curvature constant. Our Universe and its matter content can be well approximated at cosmological scales by this metric. The matter content is modeled as a perfect fluid with a given equation of state; this determines the scale factor, $a(t)$, through Einstein's equations. 

Particles at rest with the cosmic flow, with tetravelocity $u^a=\delta^a_0$, are geodesics and expands as $\Theta^{(u)}= 3 \dot{a}(t)/a(t) = 3 H(t)$, where $H(t)$ is the Hubble parameter. Despite its apparent simplicity, the FLRW solution has provoked many discussions about whether space is expanding in a real physical sense. Some misconceptions around the cosmological redshift as a Doppler effect, super-luminous velocities, the Hubble flow, the Milne universe, and other issues have been addressed in the last two decades from a general relativistic point of view (see Refs. \cite{davis} and \cite{gron}). It is now well-understood that the expansion of space is a true physical phenomenon.

If space itself is expanding, is it possible to build a rigid body? In other words, could we avoid expanding with the universe if we adopt some particular state of motion? We investigate these questions analyzing the construction of Born-rigid bodies and its quasilocal generalization in a FLRW universe. We shall proceed with our analysis separating into four cases: (i) the Newtonian approximation, (ii) the de Sitter universe, (iii) a flat FLRW, and (iv) a curved FLRW universe.

\subsection{Newtonian approximation}

The Newtonian approximation within an expanding universe is valid for proper distances below the Hubble horizon, $R \ll 1/H(t)$, where recessional velocities are subluminal. The now standard way to derive the Newtonian equations of motion is by using Fermi coordinates around an inertial observer \cite{zalda}. From this weak-gravity approximation, the geodesic equations in the Newtonian approximation are written in spherical coordinates as:

\begin{equation}
\begin{split}
\ddot{R} & = \frac{L^2}{R^3} - q(t) H(t)^2 R, \\
R^2 \dot{\phi} & = L,
\end{split}
\end{equation}
where $q(t)$ is the deceleration parameter and $L$ is the specific the angular momentum of the test particle \cite{carrera}. Note that the cosmic expansion affects test particles through its acceleration and not its kinematical expansion \cite{davis}. In this Newtonian approximation, we can then form rigid bodies introducing a force that counteracts the cosmic force simply as $\mathcal{F}_{body} = - \mathcal{F}_{\text{cosmic}}$. The energy potential of this force is $\mathcal{U}= (1/2) q(t) H(t)^2 R^2$. This simple exercise shows us that in the Newtonian approximation we can always build rigid bodies if the acceleration of the universe remains finite, i.e. if the universe does not have a Big Rip, and we consider distances below the Hubble horizon, where the approximation is valid.

\subsection{The de Sitter universe}

From the previous section, we see that if the universe is dominated by a cosmological constant, i.e. the scale factor is $a(t)=\exp(H_0t)$ with $H_0=\sqrt{\Lambda/3}$, the Newtonian cosmological force is constant, $\mathcal{F}_{\text{cosmic}}= (1/3) \Lambda R$. This force is important in studies of the turn-around radius of galaxy clusters and a potential observable of the local universe \cite{nandra}. In full GR, a constant cosmic force can be understood as a time symmetry of the general spacetime, i.e. de Sitter spacetime. 

The de Sitter metric in four dimensions can be introduced by embedding an hyperboloid with equation $X^a X_a = l^2:= 1/H_0^2$ into a $1+4$ dimensional flat spacetime with line element $ds^2= -(dX^0)^2 +(dX^1)^2+(dX^2)^2+(dX^3)^2+(dX^4)^2$. Different parametrization of the embedding describe different parts or sectors of the entire spacetime \cite{desitter}. The cosmological, or planar, coordinates of de Sitter spacetime are  given by the embedding
\begin{equation}
X_0= -l \sinh(t/l) + \frac{r^2}{2l} e^{t/l}, \quad X_1= -l \cosh(t/l) - \frac{r^2}{2l} e^{t/l}, \quad X_i = e^{t/l} x_i,
\end{equation}
with the metric
\begin{equation}
ds^2= -dt^2 + e^{2t/l} d^3x.
\end{equation}

\begin{figure}[ht!]
  \centering
  \includegraphics[width=0.5\linewidth]{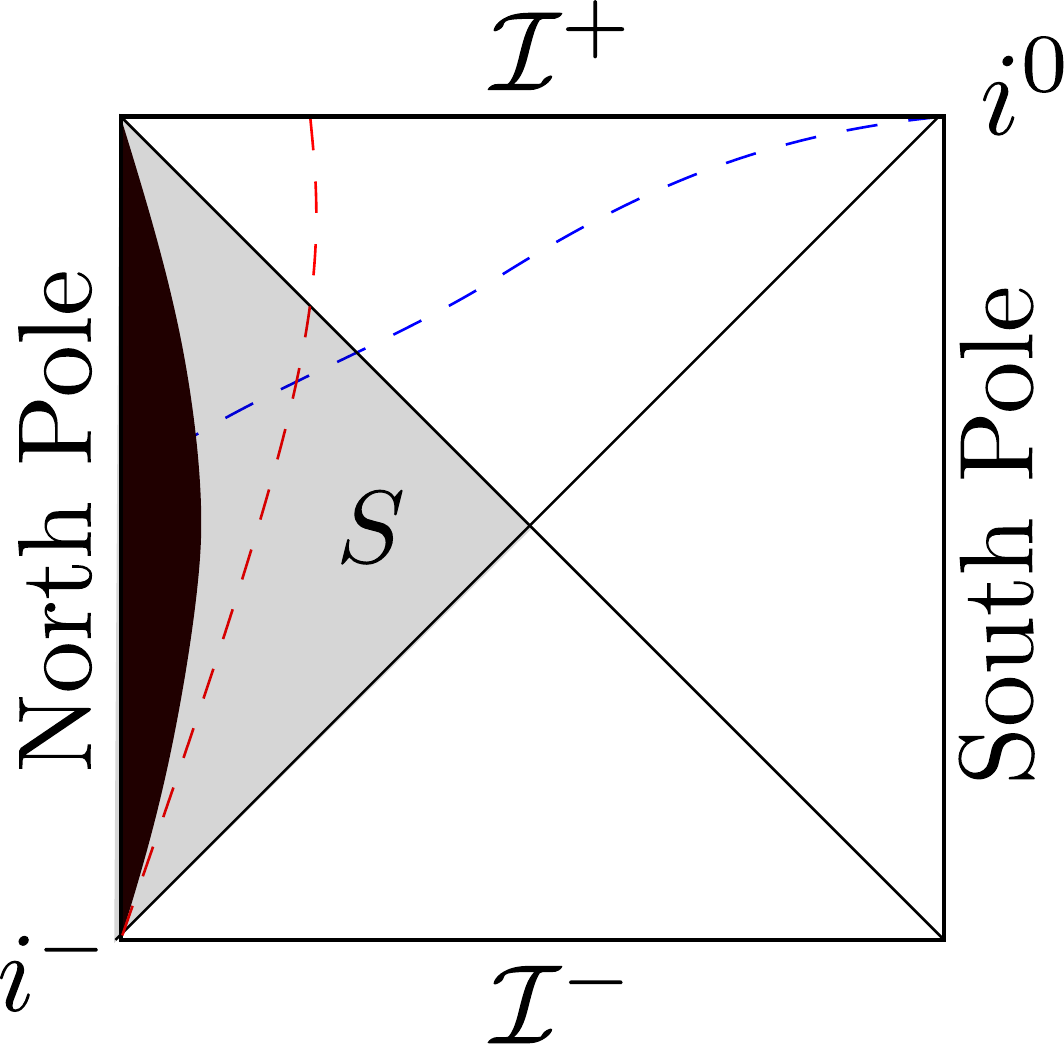}
  \caption{de Sitter conformal diagram. The gray area $S$ correspond to the static patch of the North Pole observer in the global spacetime. The black region represents a rigid body in this patch, while red and blue dashed lines correspond to the $r$ and $t$ planar coordinates, respectively}
  \label{fig: desitter}
\end{figure}

These planar coordinates describe only the causal past, $\mathcal{O}^{-}$ of an observer situated in the North pole of this spacetime (see Figure \ref{fig: desitter}) since $X_0<-X_1$. We can obtain planar coordinates for the causal future, $\mathcal{O}^{+}$, with the same embedding and transforming $X_1 \rightarrow -X_1$. The particular causal structure of de Sitter spacetime has very important consequences in deriving conservation laws and extract gravitational waves (see Ref. \cite{ashtekar}). The causal region of an observer in the North pole of this spacetime, $S=\mathcal{O}^{+} \cap \mathcal{O}^{-}$, is called the static patch.

de Sitter spacetime, being maximally symmetric, has locally ten Killing vectors, three rotations, three translations, and four boost-like vectors. Different from Minkowski, there is no global time translation Killing vector; the space-like or time-like character of all these boost Killing vectors depends on the region of the global spacetime. Focusing on the static patch of a given observer, it is possible to show that only one of this Killings vectors generates a time translation in the patch\footnote{More general, it can be shown that only a subset of four Killing vectors map the static patch into itself.}. Adapting a coordinate system to this symmetry, the metric of the static patch in these coordinates is:
\begin{equation}
ds^2 = -(1-H_0^2 R^2) dT^2 + \frac{1}{1-H_0^2 R^2} dR^2 + R^2 d\Omega^2,
\end{equation}
where $K^a=\partial^a_T$ is the Killing vector. This is analogous to building the Rindler wedge in Minkowski with a four-dimensional boost. In this way, from Theorem I, we know that a Born-rigid body can be formed following the Killing flow, with a congruence $u= K^a/||K||$, which has a radial acceleration $\alpha^a=$. Similar to the Rindler rigid body, the size of the body is constrained by $R<1/H_0$; in de Sitter, this coincides with the spatial extension of the cosmic horizon. Different from Minkowski spacetime however, the other de Sitter boosts are not time-like Killing vectors so we cannot use them to build further canonical rigid movements. This is another consequence of causal observers occupying just a portion of the whole spacetime.

Theorems presented in Section 2, well-known for Minkowski spacetime, are also valid for de Sitter spacetime, most notably the Herglotz-Noether theorem: rigid rotational motions in de Sitter are Killing motion. For instance, using the linear combination of Killing vectors $K= \partial_T + \Omega \partial_{\phi}$, this generates a rigid rotational motion as long as $K$ is time-like, i.e. if the outer radius of the body is $R < 1/H_0 \sqrt{1+\Omega/H_0}$. Note how the maximum size of the body also depends on the cosmological constant.

\subsection{FLRW flat universe}

Even though free particles are expanding in de Sitter, there is a coordinate patch adapted to a rigid congruence. This congruence has to accelerate to maintain rigidity and, contrary to Minkowski, there is a maximum size for rigidity, associated with the cosmic horizon. In the following, we will show that the true dynamic nature of the general FLRW universe makes impossible to maintain a Born-rigid body. 

If we consider comoving coordinates and a central world-line at $r=0$, geodesic distances between events at the cosmic time slice $t=0$ are given by $R= r a(t)$. Transforming to new coordinates where $R$ is the radial coordinate and $t$ remains unchanged, the metric \eqref{eq: flrwmetric} transforms as:

\begin{equation}
ds^2= -(1-H(t)^2 R^2)dt^2 - 2 H(t)R dt dR + dR^2 + R^2 d\Omega^2.
\label{eq: stationarymetric}
\end{equation}

Note that the metric depends only on the Hubble parameter, which is time dependent in this case. Stationary world-lines, with $u^a=\partial_t/||\partial_t||$, in this metric are at proper distances from the center. Hence, we might think that the congruence describes a rigid body. However, it is easy to show that the expansion $\Theta$ is non zero:
\begin{equation}
\Theta= \frac{\dot{H}(t) H(t) R^2}{\Big(1-H(t)^2 R^2\Big)^{3/2}}.
\label{eq: stationaryfried}
\end{equation} 

This can also be seen if we compute the space metric $h_{ab}$, which has an explicit time depency if $H(t)$ is not constant (i.e. if the universe is not de Sitter or Minkowksi). If we now generalize the congruence to non-stationary world-lines, we have in general a velocity field $u^a=\gamma(1,v_R(T,R),0,0)$, where we only take radial velocities because of the spherical symmetry. If we look for solutions for $\gamma$ and $v_R$ such that $\Theta_{ab}=0$, it is easy to see that the polar components of the expansion tensor are given by:
\begin{equation}
\Theta_{AB} =
\begin{pmatrix}
R^2 \gamma v_R & 0  \\
0 & R^2 \sin(\theta)^2 \gamma v_R
\end{pmatrix}.
\end{equation}
for $A,B$ = $\phi, \theta$. In order to have zero expansion, $v_R$ should be zero, i.e. stationary as in (\ref{eq: stationaryfried}). This means that a non-zero radial velocity cannot overcome the expansion. The definition of Born rigidity requires that all locally defined distances in the body should remain constant; although we can fix radial distances to the center, distances between other parts of the congruence will necessarily change when the universe is accelerated in a time-dependent way. Just as we saw with the rigid rotating system in Section 3, distances on the shell remain constant but gravity bends distances in the interior. Counter-intuitively, this cannot be counteracted by a kinematic action of the congruence, e.g. accelerating towards the center. Space itself is being stretched when $\dot{H}(t) \neq 0$. Finally, note that at linear order in $H(t)R$, the expansion is indeed zero. This is consistent with the Newtonian approximation, where particle can resist the expansion accelerating in the opposite sense, showing that this is a purely relativistic phenomenon. 

Although we cannot form rigid bodies in the bulk (i.e. Born-rigid bodies), we can always build a shell system which is non-expanding, finding $u^a$ and $s^a$ such that $\theta_{ab} \equiv 0$ (see Section 2.2). A good ansatz is of course the congruence that remain an fixed proper distance $R$, defined by the tetravelocity (\ref{eq: stationaryfried}). Now we have to find an 2-sphere with normal vector $s^a$ embedded in the congruence that does not expand. Since $s^a$ should be normal to $u^a$, in spherical symmetry this easily entails $s^a:= (0,1/\sqrt{1-H^2R^2},0,0)$. With these two vectors, we can induce the Lorentzian metric of the evolving $1+2$ screen as:
\begin{equation}
\gamma^{IJ} = 
\begin{pmatrix}
-1/(1-H^2R^2) & 0 & 0  \\
 0 & 1/R^2 & 0 \\
 0 & 0   & 1/R^2 \sin(\theta)^2  
\end{pmatrix}.
\end{equation}
and the metric on the surface as: 
\begin{equation}
\sigma_{AB} =
\begin{pmatrix}
R^2 & 0  \\
0 & R^2 \sin(\theta)^2  
\end{pmatrix}.
\end{equation}

This implies that the two-surface boundary of the congruence does not expand, $\theta_{ab}=0$. Note that the intrinsic metric of the surface is the same as a sphere in flat spacetime. 

\subsection{FLRW curved universe}

Finally, let us analyse the spatially curved FLRW universe, when $k\neq0$. Following the previous section, we change to proper distance coordinates first. In this case, these are given by \cite{faraoni}:
\begin{equation}
ds^2= -\Big( 1-\frac{H^2R^2}{1-k R^2/a^2} \Big) dt^2 -\frac{2 HR}{1-k R^2/a^2} dtdR + \frac{dR^2}{1-kR^2/a^2}+R^2 d\Omega^2.
\label{eq: flrwcurved}
\end{equation}

The stationary congruence in this spacetime with tetravelocity $u^a=\partial^a_t/||\partial_t||$, is expanding as:
\begin{equation}
\Theta= \frac{R^2 \left(k \dot{a}(t)-H(t)  \dot{H}(t) a(t)^3\right)}{g(t,R) \left(a(t)^3 \left(R^2 H(t)^2-1\right)+k R^2 a(t)\right)},
\label{eq: expansionk}
\end{equation}
where $g(t,R)=\sqrt{1-R^2 H(t)^2/(1-k R^2/a(t)^2)}$. Different from the spatially flat universe, for a suitable expansion and curvature we can build Born-rigid bodies for particular values of $k$ and $H(t)$. From (\ref{eq: expansionk}), $\Theta=0$ implies the differential equation:
\begin{equation}
\dot{H}=\frac{k}{a^2}.
\end{equation}

This equation is simply the condition of having a fixed apparent horizon in the proper distance coordinates, $dR_{\mathcal{H}}/dt=0$, where the apparent horizon is $R_{\mathcal{H}}:=1/\sqrt{H^2+k/a^2}$. For instance, for $k=-1$, one of the solutions are given by the Milne universe \cite{gron} described by $a(t)=t$, which can be mapped to a portion of the Minkowski metric; in that case, the apparent horizon is placed at infinity and there is no spatial restriction to the size of the rigid bodies. To build shell--rigid bodies we can repeat the same construction of the flat FLRW metric with the stationary congruence in the metric (\ref{eq: flrwcurved}). In the next section, we describe how we can use a rigid frame to evaluate the gravitational energy and conservation laws projected on this shell.

\begin{figure}[ht!]
  \centering
  \includegraphics[width=0.5\linewidth]{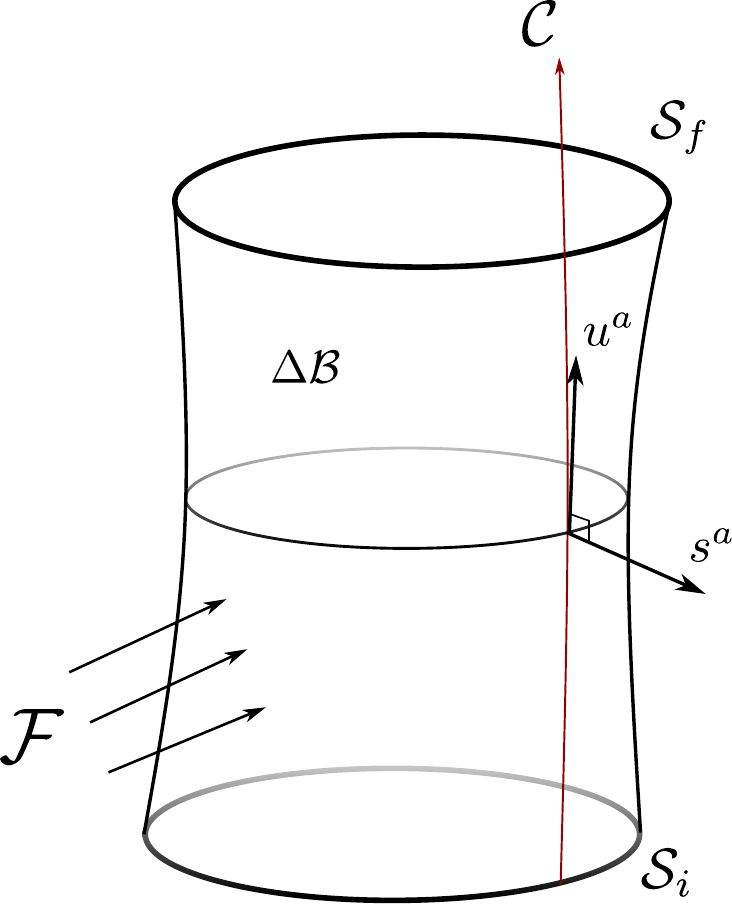}
  \caption{Representation of the quasilocal shell evolution from a state $\mathcal{S}_i$ to $\mathcal{S}_f$, formed by a congruence of world-lines $\mathcal{C}$ with time-like normal vector $u^a$ and an outward vector to the 2-sphere $s^a$. }
  \label{fig: tube}
\end{figure}

\section{Gravitational energy of the expanding universe}
\label{sec: gravenergy}

According to the Equivalence Principle, a sufficiently local body is not affected by the curvature of spacetime. This intrinsic spacetime property, different from any other matter field, is the main impediment to define a proper notion of gravitational energy. Most of the well-known constructs associated with the energy of spacetime, such as the ADM mass and the Bondi-Sachs mass \cite{poisson}, are globally defined and only suitable for asymptotically flat conditions where we can define isolated objects. In a more general context where we cannot define an isolated region, e.g. in highly dynamical spacetimes, we have to resort to quasilocal definitions of energy \cite{quasi}.

Although there are many quasilocal constructs for representing the gravitational energy, the Brown-York (BY) approach is one of the best motivated and widely used \cite{brown}. Given a spacetime region $\mathcal{D}$ with topology $\Sigma \times [t_1,t_2]$, where $\Sigma_t$ is a compact spacelike surface, we can define a time-like boundary for $\mathcal{D}$ as $\Delta \mathcal{B}=\mathcal{S} \times [t_1,t_2]$, where $\mathcal{S}$ has the topology of a two-sphere. The boundary $\Delta \mathcal{B}$ can be consider as a membrane \cite{thorne} and the two-sphere $\mathcal{S}$ as a gravitational screen \cite{freidel}. In the context of Section 2.2, the gravitational screen is the shell that evolves along the congruence. We recall that $\gamma_{ab}$ is the Lorentzian metric of the membrane $\Delta \mathcal{B}$, $s^a$ the normal vector to $\mathcal{S}$, and $u^a$ the normal time-like vector to the foliation to $\Sigma$ (see Figure \ref{fig: tube}). 

We can asociate an energy momentum tensor to the screen given by:
\begin{equation}
T^{\mathcal{B}}_{ab}= - \frac{1}{\kappa} \Big( H_{ab} -\gamma_{ab} H \Big),
\end{equation}
where $H_{ab}:= \gamma^{a'}_{a} \gamma^{b'}_{b} \nabla_{a'} s_{b'}$ is the extrinsic curvature to the membrane and $\kappa$ the coupling constant in Einstein's field equations $G_{ab} = \kappa T_{ab}$, that we set to $\kappa= 8 \pi $. This quasi-local tensor was derived by Brown and York considering the Hamilton-Jacobi formulation of the Trace K action of General Relativity. It also appears naturally as the energy-momentum tensor of time-like membranes in the Israel's junction conditions \cite{freidel}.

Using Einstein's equation and projecting it into the membrane, we obtain the holographic form of Einstein field equations that we use to obtain the conservation laws:
\begin{equation}
D_{a} T^{ab}_{\mathcal{B}}= -T^{ac} s_{a} \gamma^{b}_{c},
\label{eq: conservationby}
\end{equation}
\begin{equation}
R(\gamma) + H^{ab} H_{ab} -H^2 = -16 \pi T_{ab} s^{a} s^{b},
\end{equation}
where $D_{a}:= \gamma^{a'}_{a} \nabla_{a'}$ is the covariant derivative associated with $\gamma$, and $R(\gamma)$ is the three-dimensional Ricci scalar of the metric.

Let us consider a shell sphere evolving in time, characterized with a tetravelocity $u^a$ and spatial outward vector $s^a$, which define a membrane $\Delta \mathcal{B}$ as a closed surface $\mathcal{S}$ evolving in time. We can use this energy-momentum tensor and Einstein's field equations to compute the gravitational energy contained in the shell and its associated conservation laws. As usual, we define the quasilocal energy, momentum, and stress per area projecting onto the components of the frame:
\begin{equation}
\mathcal{E}:= u^a u^b T^{\mathcal{B}}_{ab},
\end{equation}
\begin{equation}
\mathcal{P}_a := \sigma^b_a u^c T^{\mathcal{B}}_{bc},
\end{equation}
\begin{equation}
S_{ab} = - \sigma^{c}_a \sigma^{d}_b T^{\mathcal{B}}_{cd}.
\end{equation}

Note that $\mathcal{E} \equiv - \sigma_{ab}H^{ab}/\kappa \equiv - k/\kappa$, where $k$ is the trace of the extrinsic curvature to $\mathcal{S}$. Now, we project the gravitational energy-momentum tensor with the tetravelocity $u^a$ and take the spatial derivative as:
\begin{equation}
D_a (T^{ab}_{\mathcal{B}} u_{b}) = (D_a T^{ab}_{\mathcal{B}}) u_{b} + T^{ab}_{\mathcal{B}} (D_{(a} u_{b)})
\label{eq: relationem}
\end{equation}

If we integrate $D_a (T^{ab}_{\mathcal{B}} u_{b})$ over the time-like membrane $ \Delta \mathcal{B}$ bounded by $\mathcal{S}_{1}$ and $\mathcal{S}_{2}$ we have, using Stokes theorem:
\begin{align}
\int_{\Delta \mathcal{B}}  D_a (T^{ab}_{\mathcal{B}} u_{b}) d\mathcal{B} = -\int_{\Delta \mathcal{S}} \mathcal{E}  d\mathcal{S}.
\end{align}

On the other hand, using the Gauss-Codazzi relation \eqref{eq: conservationby} we connect the spatial divergence of the gravitational energy momentum tensor with the matter flux:
\begin{equation}
(D_a T^{ab}_{\mathcal{B}}) u_{b}= - T^{ab} s_{a}u_{a}
\end{equation}

Finally, we can rewrite the last pure geometrical term in \eqref{eq: relationem} as (see Ref. \cite{epp2} for a derivation of this identity):
\begin{equation}
T^{ab}_{\mathcal{B}} \Big( D_{(a} u_{b)} \Big) = \alpha_a \mathcal{P}^a - \mathcal{S}^{ab} \theta_{ab}.
\end{equation}
where $\alpha_a$ is the acceleration projected on the space surface with the metric $\sigma_{ab}$. We these relations we obtain the balance equation:

\begin{equation}
\int_{\Delta \mathcal{S}} \mathcal{E}  d\mathcal{S} = \int_{\Delta \mathcal{B}}  \Big[ T^{ab} s_a u_b - \alpha_a \mathcal{P}^a + S^{ab} \sigma_{ab} \Big] d\mathcal{B}.
\label{eq: conservcomplete}
\end{equation}

We can rewrite Eq. (\ref{eq: conservcomplete}) simply as:
\begin{equation}
\Delta \mathcal{E} = \mathcal{F}_{\text{M}} + \mathcal{F}_{\text{ST}}.
\label{eq: conservation}
\end{equation}
where we define the conventional matter fluxes through the membrane as:
\begin{equation}
 \mathcal{F}_{\text{M}} = \int_{\Delta \mathcal{B}} T^{ab} s_a u_b d\mathcal{B},
\end{equation}
and spacetime, or geometrical fluxes as:
\begin{equation}
\mathcal{F}_{\text{ST}}= \int_{\Delta \mathcal{B}}  \Big[- \alpha_a \mathcal{P}^a + S^{ab} \sigma_{ab} \Big] d\mathcal{B}.
\end{equation}

This is the quasilocal conservation law associated with the shell $\mathcal{S}$: the change of energy inside the shell depends on the matter fluxes, $\mathcal{F}_{\text{M}}$, and spacetime fluxes, $\mathcal{F}_{\text{ST}}$ that goes through the membrane $\Delta \mathcal{B}$. The conservation law is valid for any compact surface evolving in an arbitrary spacetime. When spacetime has a preferred structure such as asymptotic-flatness or symmetries, there are conserved charges associated with them. The concept of quasilocal energy, in contrast, is valid for any spacetime. There is, however, a frame dependency. Different choices of congruences would yield different results. This is not surprising as energy is a frame-dependent concept even in Special Relativity. 

As presented for the first time in Ref. \cite{epp1}, a \textit{distinctive} frame in an arbitrary spacetime is the quasilocal rigid frame. This is, in some sense, a preferred frame since the intrinsic surface metric of the shell does not change in time as expansion tensor $\theta_{ab}$ of the shell is zero. If we apply the conservation laws to this frame, the spacetime fluxes are reduced to:
\begin{equation}
\Delta \mathcal{F}_{\rm ST} \equiv - \int_{\Delta \mathcal{B}}  \alpha_a \mathcal{P}^a  d\mathcal{B}.
\label{eq: stfluxesrigid}
\end{equation}

As it is shown in Ref. \cite{epp3}, the spacetime fluxes \eqref{eq: stfluxesrigid} have a similar form to a Poynting flux flowing through the sphere.  We will use these expressions now to evaluate the gravitational energy of a quasilocal rigid frame evolving in the expanding universe. For spherical symmetry, there is always a family of foliations where the areal radius is constant in time\footnote{This particular gauge in spherical symmetry is sometimes known as the Kodama gauge.}, as we used in previous sections (see Appendix in Ref. \cite{carrera}). Stationary observers at a fixed radius have no vorticity and no shell expansion, so the conservation law in (\ref{eq: conservation}) reduces to $\Delta \mathcal{E} = \mathcal{F}_{\text{M}}$, i.e. in spherical symmetry there is preferred frame where spacetime fluxes are zero and the total (matter plus spacetime) energy only grows or decreases by matter fluxes. This is consistent with the absence of gravitational radiation in spherical symmetry.

Although the conservation laws \eqref{eq: conservationby} are physically meaningful, there is freedom to choose a reference for the total energy at the given time \cite{brown}, which comes from the non-dynamical part of the action on the boundary. The total energy in the shell is thus:
\begin{equation}
E= \int_{\mathcal{S}} \mathcal{E}  d\mathcal{S} -\int_{\mathcal{\bar{S}}} \mathcal{\bar{E}}  d\mathcal{\bar{S}},
\end{equation}
where $ \mathcal{\bar{E}} $ is evaluated in the reference two-sphere $\mathcal{\bar{S}}$. If we want to set this energy to zero in the limit of Minkowski spacetime, the 2-sphere $\mathcal{\bar{S}}$ should be a suitably chosen manifold in flat spacetime. It is usual to take this manifold as an isometric embedding of the original two-sphere into Minkowski. For the case of spherical symmetry, in the constant radius gauge, $\mathcal{\bar{S}}$ can be chosen to be a sphere in Minkowski with extrinsic curvature $k_0= 2/R$:
\begin{equation}
E_0 = \int_{\mathcal{\bar{S}}} \mathcal{\bar{E}} = - \frac{1}{8 \pi} \int_{S^2} k_0 \: \sin(\theta) R^2 d\theta d\phi =  -R.
\end{equation}

With this reference, we can calculate the gravitational energy within a rigid shell of radius $R$ in a general FLRW universe\footnote{We would like to mention that after we sent the article for publication, related papers appeared on \texttt{ArXiv} \cite{quasicosmo,oltean2020energy} with a similar calculation of the quasilocal energy.}, using the normal vectors of Section 4.3:

\begin{equation}
E= R \Big( 1- \sqrt{1-R^2 \Big( H^2+k/a(t)^2 \Big ) }  \Big).
\label{eq: byenergy}
\end{equation}

\textit{Remarks:} The quasilocal Brown-York energy (\ref{eq: byenergy}) should be interpreted as the internal energy inside a sphere of radius $R$ \cite{brown}. Fixing $R$, we check that the system is losing or gaining energy with a rate (for $k=0$) given by:

\begin{equation}
dE/dt= R^3 H \dot{H}/\sqrt{1-H^2R^2},
\end{equation}
which is determined by the flux of matter flowing through the rigid sphere. For universes with $\dot{H}<0$ as the $\Lambda$CDM model, the system loses energy until $H(t)$ reaches $H_0$. This shows that the relevant quantity that contributes to the change in spacetime energy is the evolution of the Hubble factor and not the expansion factor itself. 

This notion of internal energy can be related in spherical symmetry with the frame independent concept of mass given by Misner-Sharp mass, which is equivalent to the Hawking mass for spherically symmetric cases. The Misner-Sharp mass is defined as $M_{MS}:= 1/2 R^3 K$, where $K$ is the sectional curvature of the $(R,t)$ plane. In this case, $M_{MS}= H^2 R^3/2 \equiv (4/3) \pi \rho R^3$. For quasilocal rigid frames, the Brown-York energy is related with the Misner-Sharp mass as \cite{visser}:
\begin{equation}
E= R \Big( 1-\sqrt{1-\frac{2 M_{MS}}{R}} \Big),
\end{equation}
\begin{equation}
M_{SH}= E- \frac{E^2}{2R}.
\end{equation}

We emphasize that this is only true for rigid frames. The $-E^2/2R$ term can be thought in general as the binding energy. Both constructs coincides in the Newtonian approximation (here, where $R \ll 1/H$) but differ in the non-linear regime. While the conservation laws we presented in spherical symmetry involves matter fluxes through the membrane of the rigid observer, the Misner-Sharp mass is associated with the flux of the conserved Kodama current, defined as $J^a:= T^{a}_{b}k^b$ where $k^a:= \epsilon^{ab} \nabla_b R$, is the Kodama vector, which is unique in spherical symmetry \cite{carrera}. The true matter flux increases the internal energy of the system, while the Kodama flux increases the mass. This could be a key factor in the thermodynamic formulation of the cosmic apparent horizon, which seems to be flawed for certain scenarios \cite{faraoni}. At the apparent horizon, the internal energy is given by $E=R_{\mathcal{H}}$ and the mass $M_{MS}=R_{\mathcal{H}}/2$. This a general characteristic of spherically symmetric spacetimes. 

Note that as we approach the apparent horizon, the quasilocal rigid frame needs a higher acceleration to keep quasilocal rigidity; at the apparent horizon, acceleration is infinite and the shell becomes a null surface. The quasilocal energy (\ref{eq: byenergy}) is thus only valid for frames where $R<R_{\mathcal{H}}$. It is interesting to compare this with the case of a black hole. We can build a quasilocal rigid frame in a Schwarzschild spacetime associated with the Killing flow of its time-like Killing vector. The frame can be extended to infinity without problems, where it reduces to the asymptotically inertial frame and the Brown-York energy approaches the ADM mass. The frame however can only maintain rigidity outside the black hole apparent horizon. If we now consider a Schwarzschild-de Sitter metric, spacetime has two apparent horizons, and the rigid frame is valid at $R_{\mathcal{H}_1}<R<R_{\mathcal{H}_2}$, where the internal energy is again $E=R_{\mathcal{H}_i}$ at both ends.

Final, if we instead take a two-sphere following the Hubble flow, the internal Brown-York energy do not receive an influx of matter (i.e. we are comoving with the matter flow). Since the system is not quasirigid, the spacetime flux in the third term of the conservation equation is not zero.  Using a non-rigid embedding into Minkowski spacetime, however, the Brown-York energy, as calculated by Ref. \cite{afshar}, is exactly zero.

\section{Conclusions}

We have taken a fresh look into rigid systems using the century-old Born-rigid concept and the newly introduced quasilocal-rigid concept. We reviewed the main theorems regarding Born-rigid bodies and we presented a clean way to build rigid and quasilocal rigid bodies in flat spacetime based on a frame that uses space geodesics as coordinates. With these results, we investigated the notion of rigidity in the expanding universe, i.e. the possibility of building a non-expanding body in the relativistic sense. We have shown that this is impossible in a general FLRW universe except in de Sitter spacetime and in the Newtonian approximation. For a dynamical FLRW universe, we can construct only quasilocal rigid bodies, i.e. non-expanding shells, using accelerated observers. As we explained in detail, these are adequate frames to analyze the gravitational energy contained in a compact region of the expanding universe. We showed that the energy is not zero, contrary to previous results, and it is determined by the flux of cosmic dust as seen by the rigid frame.

\begin{acknowledgements}
We thank the reviewer for valuable suggestions that improved the quality of the paper.
This work was supported by the Argentine agency CONICET
(PIP 2014-00338), the National Agency for Scientific
and Technological Promotion (PICT 2017-0898), and
the Spanish Ministerio de Econom\'ia y Competitividad
(MINECO/FEDER, UE) under grant AYA2016-76012-C3-1-P.
\end{acknowledgements}

%
%



\bibliographystyle{spphys}       

\bibliography{bibliography.bib}   

\end{document}